\documentclass[%
 reprint,
 amsmath,amssymb,
 aps,
]{revtex4-2}

\usepackage{graphicx}
\usepackage{dcolumn}
\usepackage{hyperref}
\usepackage{xcolor}
\usepackage{mathtools}
\usepackage{subcaption}
\usepackage[lined, ruled]{algorithm2e}

\begin{document}

\preprint{APS/123-QED}

\title{A variational quantum eigensolver tailored to multi-band tight-binding simulations of electronic structures}

\author{Dongkeun Lee}
\affiliation{Center for Quantum Information R\&D, Korea Institute of Science and Technology Information, Daejeon 34141, Republic of Korea}

\author{Hoon Ryu}
\email{elec1020@kumoh.ac.kr}
\affiliation{School of Computer Engineering, Kumoh National Institute of Technology, Gumi, Gyeongsangbuk-do 39177, Republic of Korea}%

\date{\today}

\begin{abstract}
We propose a cost-efficient measurement scheme of the variational quantum eigensolver (VQE) for atomistic simulations of electronic structures based on a tight-binding (TB) theory. Leveraging the lattice geometry of a material domain, the
sparse TB Hamiltonian is constructed in a bottom-up manner and is represented as a linear combination of the standard-basis (SB) operators. The cost function is evaluated with an extended version of the Bell measurement circuit that can
simultaneously measure multiple SB operators and therefore reduces the number of circuits required by the evaluation process. The proposed VQE scheme is applied to find band-gap energies of metal-halide-perovskite supercells that have
finite dimensions with closed boundaries and are described with a $sp^3$ TB model. Experimental results confirm that the proposed scheme gives solutions that follow well the accurate ones, but, more importantly, has the computing efficiency
that is obviously superior to the commutativity-based Pauli grouping methods. Extending the application scope of VQE to three-dimensional confined atomic structures, this work can serve as a practical guideline for handling TB simulations in the
noise-intermediate-scale quantum devices.\end{abstract}
\maketitle


\section{Introduction}

Quantum computers have been expected to efficiently simulate diverse many-body quantum systems, having the potential to surpass the performance of classical computers \cite{Feynman.1982}. It was first proposed that the eigenvalue problem
of a given Hamiltonian can be tackled with a universal quantum computer based on the quantum phase estimation (QPE) algorithm \cite{Kitaev.1995}. Though the QPE algorithm shows an exponential speedup that is not easy to be secured with
classical computers \cite{Abrams.1997, Abrams.1999}, it requires many controlled unitary gates to solve large-scale systems so the fault-tolerant quantum computer is essential. Despite of ongoing significant development in quantum hardware
and software, QPE is not yet feasible in near-term quantum computers that lie in the noisy intermediate-scale quantum (NISQ) regime \cite{Preskill.2018}, motivating strong needs for quantum-classical hybrid algorithms for leveraging NISQ devices.
Ever since its initial proposal for tackling small-size molecular Hamiltonians \cite{Peruzzo.2014}, the variational quantum eigensolver (VQE) has been the most extensively utilized and developed algorithms over the past decade for simulations
of electronic structures in NISQ computers with support of classical computing \cite{vqa, Tilly.2022, Bharti.2022}. Although the VQE algorithm has challenges such as barren plateaus \cite{McClean.2018} and NP-hardness of the classical optimization
process \cite{Bittel.2021}, alternative or breakthrough strategies tailored to specific conditions have been devised to circumvent these difficulties \cite{Larocca.2024}, so VQE still belongs to practical applications that can exploit NISQ computers.

Many studies on quantum algorithms involving VQE and QPE so far have predominantly investigated properties of molecular systems or lattice models by encoding qubits with second quantization \cite{Tilly.2022, Bharti.2022}. However, application
studies based on encoding with first quantization still remain relatively rare in the context of VQE \cite{Tilly.2022}, though, from a resource perspective, the use of first quantization can be beneficial for computing electron systems in certain cases
\cite{Abrams.1999, Berry.2018}. A good candidate for computation based on first quantization is the empirical simulation of nanoscale materials \cite{Wang2004, Sengupta2015, Vogl1983, Biswas1994, Jancu1998, Richard2016}, whose focal
characteristic is to describe a single-electron Hamiltonian with a fixed set of parameters. In particular, simulations with the atomistic tight-binding (TB) model have been extensively employed to study electrical properties of realistically sized solid
crystalline structures that is in principle hard to be done with $ab$-$initio$ simulations, such as wavelength engineering of quantum dot devices \cite{Eric2024, Ryu2019}, transport characteristics of nanowire transistors \cite{Huang2018, Prentki2020},
and designs of novel electronic devices like atomic-scale interconnects or single electron transistors \cite{Ryu2015, Weber2014, Ryu2014}.

Recently, TB simulations have also gained attention as a sound application of VQE \cite{Cerasoli.2020, Sherbert.2021, Sherbert.2022}. Two previous works focus on the calculation of single-electron TB Hamiltonians with second quantization that
eliminates the need for intricate fermion-to-Pauli mappings such as the Jordan-Wigner and the Bravyi-Kitaev transformations \cite{Sherbert.2021, Sherbert.2022}. They however require $O(D)$ qubits to tackle Hamiltonians of dimension $D$ (= the
size of single-particle basis functions) so challenge in scalability could exist in NISQ computers. Cerasoli $et$ $al$. \cite{Cerasoli.2020} computed the Silicon bulk bandstructure with $O(\log D)$ qubits, by representing the TB Hamiltonian with first
quantization. But, the target problem is rather small (a single primitive unit cell consisting of two atoms), and, more importantly, the Hamiltonian is mapped with the most primitive Pauli decomposition that can in principle require the computation of
up to 4$^{O(\log D)}$ coefficients.

So in this work, we propose a new variant of VQE, focusing on cost-efficient handling of TB Hamiltonians with a set of the first-quantized atomic orbitals. By leveraging the geometric information of three-dimensional (3D) finite-size supercells, we
first show that the bottom-up construction of the Hamiltonian matrix can be done with a set of standard basis (SB) operators that is known to be useful for description of any vectors and matrices \cite{Haley.1972}. Then, we discuss a measurement
scheme that computes the cost function designed to determine the band-gap energy of a confined supercell. To reduce the number of quantum circuits required to evaluate the cost function, we group SB operators obtained with decomposition of
the TB Hamiltonian and employ the Greenberger-Horne-Zeilinger (GHZ) measurement that is a generalized version of the Bell measurement \cite{Liu.2021, Kondo2022, Ali.2023}. Results reveal that our measurement scheme remarkably outperforms
Pauli-based approaches because the cost of SB decomposition \& grouping is linearly proportional to the number of nonzero elements in the Hamiltonian matrix, and our measurement circuit, requiring at most $N$ CNOT gates ($N$ = the number
of qubits in the circuit), can be more concise than the one that simultaneously measures commuting Pauli strings \cite{Crawford.2021}. Finally, we verify the proposed method by conducting numerical simulations to find the band-gap energy of 3D
metal-halide-perovskite supercells that are described with a $sp^3$ TB model \cite{Richard2016, Ryu2019}.

\section{Methods}\label{sec:methods}

\subsection{TB Model and Hamiltonian}\label{sec:methods_tbh}

The TB model is one of theoretical frameworks used to describe electronic structures of crystalline materials. It is based on the assumptions that a single electron occupies localized atomic orbitals on individual atoms residing in a material structure
and that the overlap of atomic orbitals occurs only between nearest neighboring atoms. The TB Hamiltonian in the first-quantized basis $\{|n,i\rangle\}$, where $n$ and $i$ indicate the orbital index and atomic site, respectively, can be written as

\begin{equation}\label{eq:tbh}
\hat{\mathcal{H}} = \sum_{n',n,i',i} t_{n'n}^{i'i}|n',i'\rangle\langle n,i|,
\end{equation}
where $t_{n'n}^{i'i}=\langle n',i'|\hat{\mathcal{H}}|n,i\rangle$ is the single-electron integral involving the on-site energy of each atoms and the overlap energies between two nearest atoms. The element $t_{n'n}^{i'i}$ is then given as

\begin{align}\label{eq:integ}
t^{i'i}_{n'n} = \int d\mathbf{r}\;\psi^{\ast}_{n'}(\mathbf{r},\mathbf{R}_{i'})\hat{\mathcal{H}}(\mathbf{r})\psi_n(\mathbf{r},\mathbf{R}_i),
\end{align} 
where $\psi_n(\mathbf{r},\mathbf{R}_i)$ denotes the basis function that describes a single electron in $n$th localized atomic orbital with the lattice vector $\mathbf{R}_i$. Since atoms are assumed to be tightly coupled, the overlap integral $t^{i'i}_{n'n}$
is zero when $i'$ $\neq$ $i$ except the case that $i'$th atom belongs to nearest neighbors of the $i$th atom, and therefore $\hat{\mathcal{H}}$ becomes a sparse matrix. Due to the translational symmetry of crystalline solid atomic structures, $\mathbf{R}_i$
for all atomic indices $i$ in the lattice can be spanned with the primitive translation vectors. Usually, the $t^{i'i}_{n'n}$ terms of the TB Hamiltonian in Eqs.~\eqref{eq:tbh} and (\ref{eq:integ}) are empirically secured by fitting the bulk bandstructure that
is known by either experiments or $ab$-$initio$ calculations. Since the on-site and coupling overlap integrals are dictated by atomic types and positions, we can easily identify nonzero elements and their row \& column indices in the TB Hamiltonian
by examining the types and primitive vectors of all the atoms from the geometric information (atomic species and positions) of the supercell composed of a given material.

\subsection{Measurement Scheme for Expectation Value}\label{sec:methods_sbghz}

To determine eigenpairs of a given matrix, VQE needs the expectation value of the observable that involves the target matrix. To evaluate the expectation value with quantum circuits, the observable needs to be decomposed into a weighted sum of
operators that can be readily manipulated with gate operations. Hamiltonians describing electronic structures based on a set of orthogonal basis are always hermitian, and can be typically expressed as a linear combination of Pauli operators with approaches
like the Jordan-Wigner transformation \cite{Jordan.1928}. As the Hamiltonian becomes larger in size, however, the number of associated Pauli strings rapidly increase \cite{Hantzko.2024}, so strategies for grouping Pauli strings have been adopted to
reduce the effective number of Pauli strings with the general commutativity (GC), qubit-wise commutativity (QWC), and unitary partitioning \cite{McClean.2016,Gokhale.2019,Gokhale.2019utu,Yen.2020,Verteletskyi.2020dt,Kandala.2017,Izmaylov.2019}.
A cost-efficient circuit that can simultaneously measure $k$ commuting Pauli strings with up to $kN$ CNOT gates, has been also reported \cite{Crawford.2021}. In this subsection, we present the methodological details of our measurement scheme
and grouping strategy that are well-suited for sparse TB Hamiltonian matrices with better performance than existing approaches. 

\subsubsection{Representation of TB Hamiltonian with SB Operators}
The SB in linear algebra, a set of $L$-dimensional vectors having a single nonzero element of 1, can span the entire vector space $\mathbb{R}^L$ or $\mathbb{C}^L$. The computational basis $\{|\mathbf{z}\rangle\}$ for $N$
qubits in quantum computing is nothing but the SB of the pure quantum state, where $\mathbf{z}\equiv z_1\cdots z_N$ for each $z_j\in\{0,1\}$. Adopting this basis in the quantum circuit, we can define a SB operator $\{|\mathbf{z}\rangle\langle\mathbf{z}'|\}$
whose functionality is to map one computational basis state to another. Notably, any $N$-qubit SB operator can be factorized as a tensor product of single-qubit SB operators, which is given by 
\begin{equation}\label{eq:std_basisop}
    |\mathbf{z}\rangle\langle\mathbf{z}'|=\bigotimes_{j=1}^{N}|z_j\rangle\langle z'_j|, 
\end{equation}
where $z_j$ denotes the binary value of the $j$th qubit and $|z_j\rangle\langle z'_j|$ represents the SB for single-qubit operators at the $j$th site. These SB operators can obviously facilitate the representation of an arbitrary observable $\hat{\mathcal{O}}$
as follows,
\begin{align}\label{eq:std_decomp}
\hat{\mathcal{O}} = \sum_{\mathbf{z},\mathbf{z}'} o_{\mathbf{z},\mathbf{z}'}|\mathbf{z}\rangle\langle\mathbf{z}'|,
\end{align}
where $o_{\mathbf{z},\mathbf{z}'}$'s are complex coefficients. Representation of sparse TB Hamiltonians with Eq. \eqref{eq:std_decomp} is then straightforward since the row and column index of a specific nonzero element can be directly mapped to bit
strings $\mathbf{z}$ and $\mathbf{z}'$ by converting indices to binary values, and the coefficient $o_{\mathbf{z},\mathbf{z}'}$ indicates the corresponding nonzero value. Following this scheme, we only need to map up to $M$ terms if a given Hamiltonian
has $M$ nonzero elements. Consequently, in general, the SB-based strategy in Eqs. \eqref{eq:std_basisop} and \eqref{eq:std_decomp} can be much more efficient in terms of computing cost than the Pauli decomposition whose cost grows exponentially
with the number of qubits needed to describe the target Hamiltonian \cite{Hantzko.2024}.

\subsubsection{GHZ Measurement Circuit}

\begin{figure}[!t]
    \centering
    \includegraphics[width=\columnwidth]{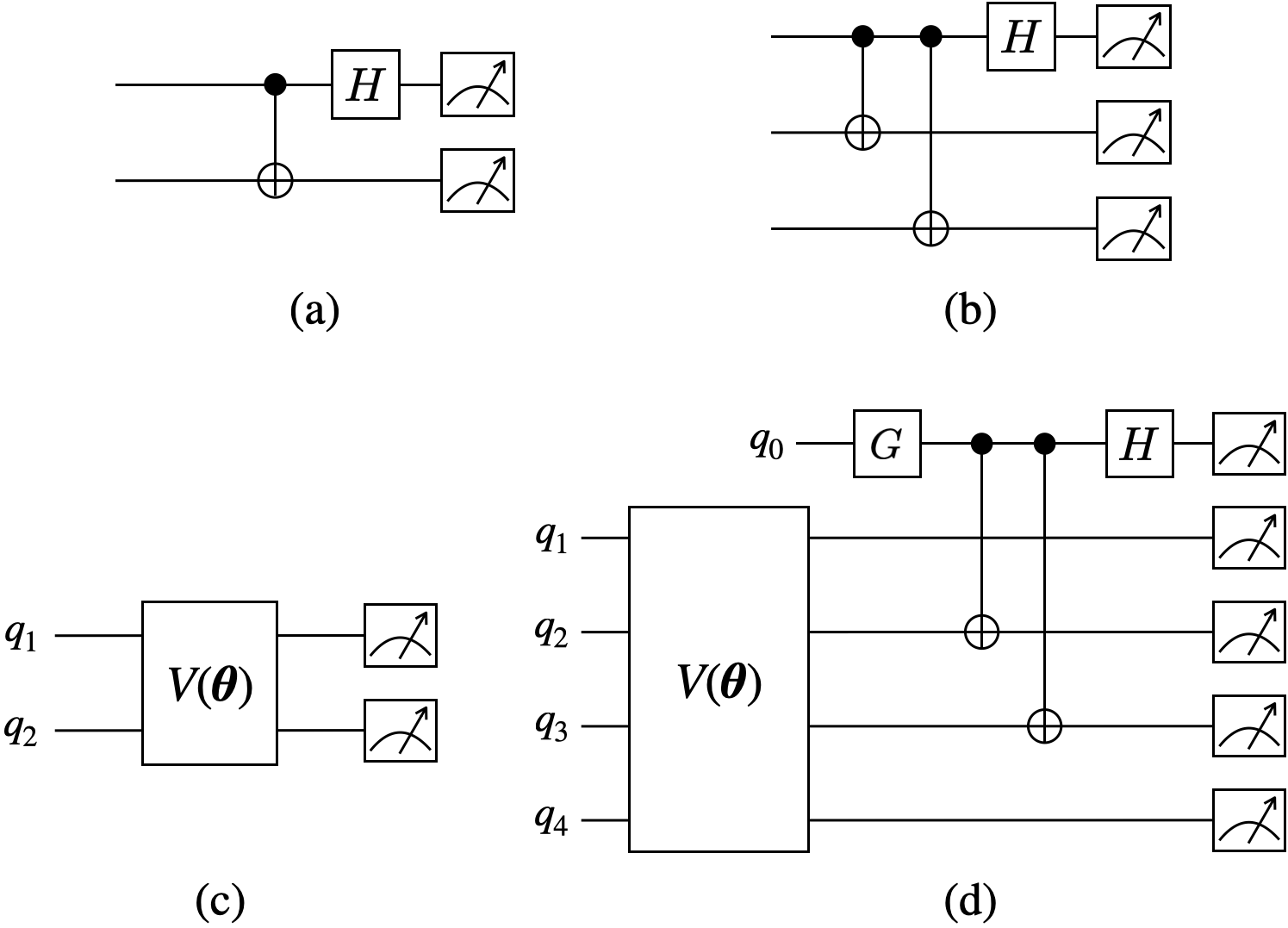}
    \caption{\small Circuits for the (a) Bell measurement and (b) three qubit GHZ measurement, and circuits for estimation of (c) $\langle\psi|z_1z_2\rangle\langle z_1z_2|\psi\rangle$ and
    (d) $\langle\psi|z_1z_2z_3z_4\rangle\langle z_1(z_2\oplus1)(z_3\oplus1)z_4|\psi\rangle$. Non-Hermitian operators can be measured by applying the GHZ measurement to the tensor product of $G|0\rangle$ and
    $|\psi\rangle=V(\boldsymbol{\theta})|\mathbf{0}\rangle$, where $G\in\{H, S\cdot H\}$ and $V(\boldsymbol{\theta})$ is an ansatz circuit.}
    \label{fig:circ_diagrams}
\end{figure}

The Bell measurement, being also known as the Bell state measurement, is a joint measurement scheme that projects two qubits onto one of the four Bell states: $\{|00\rangle\pm|11\rangle, |01\rangle\pm|10\rangle\}$. In a quantum circuit, the Bell measurement
can be implemented using a sequence of a CNOT gate, a Hadamard ($H$) gate, and $Z$-basis measurements, as shown in Fig.~\ref{fig:circ_diagrams}(a). The Bell measurement can be extended to multi-qubit ($>$ 2) joint measurements, analogous to the
way that Bell states are generalized to multi-qubit GHZ states that have the genuine multipartite entanglement. The sequence of quantum gates needed to implement a circuit for the GHZ measurement can be expressed as follows,
\begin{equation}\label{eq:U_Bell}
    U_{GHZ} = H_0\left(\prod_{j\in \mathcal{X}} CX_{0,j}\right),
\end{equation}
where $CX_{0,j}$ represents the CNOT gate with the (control, target) qubit at the ($0$th, $j$th) site, and $\mathcal{X}$ denotes the set of qubits employed as target qubits of CNOT gates. Conducting $Z$-basis measurements at the end of an $N$-qubit $U_{GHZ}$ 
yields one of $2^N$ different binary strings that determines the type of projection driven by the entire circuit ($U_{GHZ}$ + $Z$-basis measurement). In the three qubit circuit shown in Fig.~\ref{fig:circ_diagrams}(b) as an example, an outcome of 010 leads to the
GHZ measurement being performed as $U_{GHZ}|010\rangle\langle 010|U^\dagger_{GHZ} = (|010\rangle+|101\rangle)(\langle 010|+\langle 101|)/2$.

\subsubsection{Expectation Value of SB Operator}

The GHZ measurement can be used to evaluate the expectation value of a SB operator with an $N$-qubit ansatz $V(\boldsymbol{\theta})$ and a state $|\psi\rangle\equiv V(\boldsymbol{\theta})|\mathbf{0}\rangle$. According to Eq. \eqref{eq:std_basisop}, it is
easy to get the expectation value of SB operators that consist solely of projective measurements $|z_j\rangle\langle z_j|$ on individual qubits, as depicted in Fig.~\ref{fig:circ_diagrams}(c) that indicates the expectation value of the SB operator
$|\mathbf{z}\rangle\langle\mathbf{z}|$ is equal to the probability $p(\mathbf{z})$ of the outcome $\mathbf{z}$.

If non-Hermitian operators $|z_j\rangle\langle z_j\oplus 1|$ act on certain qubits in a given SB operator, the expectation value of the operator can be computed with an additional ancilla qubit followed by a single-qubit logic $G\in\{H, S\cdot H\}$ (a single $H$ gate or
a sequence of the $H$ \& $S$ gate) as shown in Fig.~\ref{fig:circ_diagrams}(d). The GHZ measurement is then applied to this ($N$+1)-qubit circuit where the ancilla qubit acts as a control qubit for up to $N$ CNOT gates in $U_{GHZ}$ and some of the remaining
$N$ qubits are used as target qubits, belonging to the set $\mathcal{X}$ defined in Eq. \eqref{eq:U_Bell}. Here, the set $\mathcal{X}$ is determined by the SB operator whose expectation value is to be estimated:
$\mathcal{X} \equiv \{ j | x_j=1 \text{ for } \mathbf{x}, 1\le j \le N \}$ for a given SB operator $| \mathbf{z} \rangle\langle \mathbf{z}'|$, where $\mathbf{z}' \equiv \mathbf{z}\oplus\mathbf{x}$ with a bit string $\mathbf{x}\equiv x_1\cdots x_N$. If a SB operator is
$|0101\rangle\langle 0011|$, for example, $\mathbf{x}=0110$ contributes to the set $\mathcal{X}=\{2,3\}$ so $U_{GHZ}$ has two CNOT gates (see Fig.~\ref{fig:circ_diagrams}(d)). Performing the GHZ measurement on this ($N$+1)-qubit circuit where the ancilla
qubit is the 0th qubit, we get the probabilities that the ($N$+1)-bit outcome is $0z_1\cdots z_N$ (0$\mathbf{z}$) or $1z_1\cdots z_N$ (1$\mathbf{z}$). These probabilities are used to obtain the expectation value of the SB operator
$|\mathbf{z} \rangle\langle \mathbf{z}'|$ as follows,
\begin{align}\label{eq:expval_stdop}
    \langle\psi|\mathbf{z} \rangle\langle \mathbf{z}' |\psi\rangle  & = [p_H(0\mathbf{z}) - p_H(1\mathbf{z})]\\
    \nonumber &\;\;\;\;\; - i[p_{SH}(0\mathbf{z})-p_{SH}(1\mathbf{z})],
\end{align}
where $p_H$ and $p_{SH}$ are the probabilities obtained with $G$ = $H$ and $S\cdot H$, respectively. The difference between $p_G(0\mathbf{z})$ and $p_G(1\mathbf{z})$ can be evaluated with a single circuit as follows,
\begin{align}\label{eq:prob_diff}
    p_G(0\mathbf{z})-p_G(1\mathbf{z}) &= \langle G|0\rangle\langle 1|G\rangle\cdot\langle\psi|\mathbf{z} \rangle\langle \mathbf{z}'|\psi\rangle \\
    \nonumber &\;\;\;\;\; + \langle G|1\rangle\langle 0|G\rangle\cdot\langle\psi|\mathbf{z}' \rangle\langle \mathbf{z}|\psi\rangle,
\end{align}
where $|G\rangle = G|0\rangle$. We note that the detailed derivation of Eqs. \eqref{eq:expval_stdop} and \eqref{eq:prob_diff} is provided in Appendix~\ref{sec:derv1}.
 
\subsubsection{Grouping SB Operators}\label{sec:grouping}

Eqs. \eqref{eq:std_decomp} and \eqref{eq:expval_stdop} clearly indicate the expectation value of a TB Hamiltonian having $M$ nonzero elements can be estimated with $O(2M)$ circuits. This number of quantum circuits, however, can be further reduced with a
proper partition of the $M$ SB operators that can be achieved by means of the GHZ measurement circuit. Let us consider a set of ($N$+1)-qubit circuits that consist of an $N$-qubit ansatz, a single-qubit $G$ logic on the ancilla qubit, and a GHZ measurement,
as illustrated in Fig. \ref{fig:circ_diagrams}(d). Then, as Eq.~\eqref{eq:expval_stdop} indicates, all the possible outcomes $\mathbf{z}$ with probability $p(\mathbf{z})$ can be leveraged to get the expectation value of corresponding SB operators
$|\mathbf{z}\rangle\langle \mathbf{z}\oplus\mathbf{x}|$, provided that the GHZ measurement is performed with a set of $\mathcal{X}$ that is determined by $\mathbf{x}$. In other words, a single GHZ measurement dedicated to a specific $\mathcal{X}$ can
$simultaneously$ measure SB operators $|\mathbf{z}\rangle\langle \mathbf{z} \oplus \mathbf{x}|$ for all $\mathbf{z}$, where $x_j=1$ for $j\in\mathcal{X}$ and $x_j=0$ otherwise. Consequently, the expectation value of a TB Hamiltonian can be efficiently
estimated by grouping associated SB operators based on $\mathbf{x}$, which reduces the number of required circuits to the number of distinct $\mathbf{x}$'s. The detailed procedure for grouping SB operators is presented in Algorithm~\ref{alg}.

\begin{algorithm}[t]
\SetKwData{Left}{left}\SetKwData{This}{this}\SetKwData{Up}{up}
\SetKwFunction{Union}{Union}\SetKwFunction{FindCompress}{FindCompress}
\SetKwInOut{Input}{input}\SetKwInOut{Output}{output}
\Input{$M$ nonzero coefficients $o_{\mathbf{z},\mathbf{z}'}$ and pairs of binary indices $\mathbf{z}$, $\mathbf{z}'$ ($\mathbf{z}\le\mathbf{z}'$)}
\Output{$\mathcal{G}^R_{\mathbf{x}}$ and $\mathcal{G}^I_{\mathbf{x}}$ $\forall \mathbf{x}$}
\BlankLine
\ForAll{nonzero $o_{\mathbf{z},\mathbf{z}'}$}{
$\mathbf{x} \leftarrow \mathbf{z}\oplus\mathbf{z}'$\\
$\mathcal{X} \leftarrow \{ j | x_j=1 \text{ for } \mathbf{x}, 1\le j \le N \}$\\
\uIf{$\rm{Re}[o_{\mathbf{z},\mathbf{z}'}]\neq0$}{
\lIf{$\mathcal{G}^R_{\mathbf{x}}$ does not exist}{$\mathcal{G}^R_{\mathbf{x}}=\emptyset$}
$\mathcal{G}^R_{\mathbf{x}}\leftarrow \mathcal{G}^R_{\mathbf{x}}\cup \{\mathbf{z}\}$
}
\ElseIf{$\rm{Im}[o_{\mathbf{z},\mathbf{z}'}]\neq0$}{
\lIf{$\mathcal{G}^I_{\mathbf{x}}$ does not exist}{$\mathcal{G}^I_{\mathbf{x}}=\emptyset$}
$\mathcal{G}^I_{\mathbf{x}}\leftarrow \mathcal{G}^I_{\mathbf{x}}\cup \{\mathbf{z}\}$
}
}
\caption{Grouping SB operators based on the GHZ measurement for estimation of $\langle\hat{\mathcal{O}}\rangle$}\label{alg}
\end{algorithm}

Two features of TB Hamiltonians can make the grouping procedure in Algorithm~\ref{alg} be more efficient. First, due to their Hermiticity, all the Hermitian conjugate pairs $|\mathbf{z}\rangle\langle \mathbf{z}'|$ and $|\mathbf{z}'\rangle\langle\mathbf{z}|$ can
be grouped, and the search space \{$\mathbf{z}$, $\mathbf{z}'$\} in Algorithm 1 can be further narrowed down to the cases that satisfy the condition $\mathbf{z}\le\mathbf{z}'$. Second, not all the elements in TB Hamiltonians have both real and imaginary parts.  
So, the number of circuits required to estimate the expectation value can be reduced by half for matrix elements ($o_{\mathbf{z},\mathbf{z}'}$) that are purely real or imaginary, as can be supported by Eqs. \eqref{eq:std_decomp} and \eqref{eq:expval_stdop}. 

Our strategy for grouping SB operators with the GHZ measurement has non-trivial advantages over Pauli-based methods. First, SB grouping does not require optimization processes based on the commutativity of operators that are essential for Pauli grouping,
and incurs only an $O(M)$ cost ($M$ = the number of nonzero elements). Second, the measurement circuit for simultaneously estimating $k$ commutating Pauli strings requires roughly at most $O(kN/\log k)$ CNOT gates \cite{Crawford.2021}. The GHZ
measurement circuit, however, can evaluate the expectation value of SB operators with a single $H$ and at most $N$ CNOT gates.

Finally, we note that Algorithm \ref{alg} is specialized for TB simulations upon the work reported by Kondo $et$ $al.$ \cite{Kondo2022}. Compared to the original appraoch, ours treats nonzero elements with the condition $\mathbf{z}\le\mathbf{z}'$ to handle
Hermitian matrices with lower cost, and can additionally reduce the number of required circuits by up to half depending on the type of elements of TB Hamiltonians.

\subsection{Optimal Number of Shots}\label{sec:samp}

Due to the probabilistic nature of quantum systems, the expectation value of any observables needs to be evaluated with a set of samples that are obtained by repeated measurements, so the error driven by finite sample sizes (sampling errors) inevitably
affects the preciseness of any quantum algorithms. The sampling error has been analyzed in the context of VQE where the Hamiltonian is expressed as a linear combination of Pauli strings \cite{Wecker.2015,Rubin.2018}, and, in the same vain, here we
discuss the sampling error involved in estimating the expectation of an observable $\hat{\mathcal{O}}$ in Eq.~\eqref{eq:std_decomp} with our grouping method presented in Algorithm~\ref{alg}. 

The sampling error $\epsilon$ stemming from the estimation of $\langle\hat{\mathcal{O}}\rangle$ is the statistical error that arises due to the nonzero variance of samples, $\epsilon^2\equiv\text{Var}(\langle\hat{\mathcal{O}}\rangle)$. Since its $N$-qubit SB
operators are grouped based on the bit string $\mathbf{x}$ and the type of the coefficient $o_{\mathbf{z},\mathbf{z}'}$, the observable $\hat{\mathcal{O}}$ in Eq.~\eqref{eq:std_decomp} can be reorganized as follows,
\begin{align}
    \label{eq:obs_ghzm}\hat{\mathcal{O}} = \sum_{\mathbf{x}}\sum_{\mathbf{z}} &\text{Re}[o_{\mathbf{z},\mathbf{z}\oplus\mathbf{x}}]\cdot |\mathbf{z}\rangle\langle\mathbf{z}\oplus\mathbf{x}|\\
    \nonumber & +\sum_{\mathbf{x}\neq 0}\sum_{\mathbf{z}} \text{Im}[o_{\mathbf{z},\mathbf{z}\oplus\mathbf{x}}]\cdot i|\mathbf{z}\rangle\langle\mathbf{z}\oplus\mathbf{x}|.
\end{align}
Each term in Eq.~\eqref{eq:obs_ghzm} corresponds to a single GHZ measurement for each bit string $\mathbf{x}$ that is performed on the state $|\psi\rangle|G\rangle$ with $G=H$ and $S\cdot H$ for the real and imaginary part of the coefficient, respectively. 
By applying the formula for the sample variance ($\text{Var}(\langle\hat{A}\rangle)=\text{Var}(\hat{A})/n$ with $n$ shots) to Eq.~\eqref{eq:obs_ghzm}, the variance of the sample mean for $\langle\hat{\mathcal{O}}\rangle$ can be expressed as follows,
\begin{align}\label{eq:samp_err}
    \nonumber \epsilon^2 =  \sum_{\mathbf{x}}& \frac{\text{Var}(\sum_{\mathbf{z}}\text{Re}[o_{\mathbf{z},\mathbf{z}\oplus\mathbf{x}}]|\mathbf{z}\rangle\langle\mathbf{z}\oplus\mathbf{x}|)}{m_{R,\mathbf{x}}} \\
    & + \sum_{\mathbf{x}\neq 0} \frac{\text{Var}(\sum_{\mathbf{z}}i\text{Im}[o_{\mathbf{z},\mathbf{z}\oplus\mathbf{x}}]|\mathbf{z}\rangle\langle\mathbf{z}\oplus\mathbf{x}|)}{m_{I,\mathbf{x}}}, 
\end{align}
where $m_{R(I),\mathbf{x}}$ denotes the number of shots obtained from the repeated GHZ measurements that are associated with the bit string $\mathbf{x}$ and the real(imaginary) part of $o_{\mathbf{z},\mathbf{z}\oplus\mathbf{x}}$. With Lagrange
multiplier, the total number of samples ($m=\sum_{\mathbf{x}} m_{R,\mathbf{x}} + \sum_{\mathbf{x}\neq 0} m_{I,\mathbf{x}}$) can be minimized with the constraint of Eq.~\eqref{eq:samp_err}. The upper bound of this minimum sample number is given as
\begin{align}\label{eq:samp_err_bound}
\frac{1}{\epsilon^2} &\left( \sum_{\mathbf{x}} \max_{\mathbf{z}}|\text{Re}[o_{\mathbf{z},\mathbf{z}\oplus\mathbf{x}}]| + \sum_{\mathbf{x}\neq 0} \max_{\mathbf{z}} |\text{Im}[o_{\mathbf{z},\mathbf{z}\oplus\mathbf{x}}]| \right)^2,
\end{align}
which can roughly tell us how many shots are required to estimate $\langle\hat{\mathcal{O}}\rangle$ with error $\epsilon$ using our grouping method. We note that the detailed derivation of Eq.~\eqref{eq:samp_err_bound} is provided in Appendix~\ref{sec:derv2}.

\begin{figure}[t]
    \centering
    \includegraphics[width=\columnwidth]{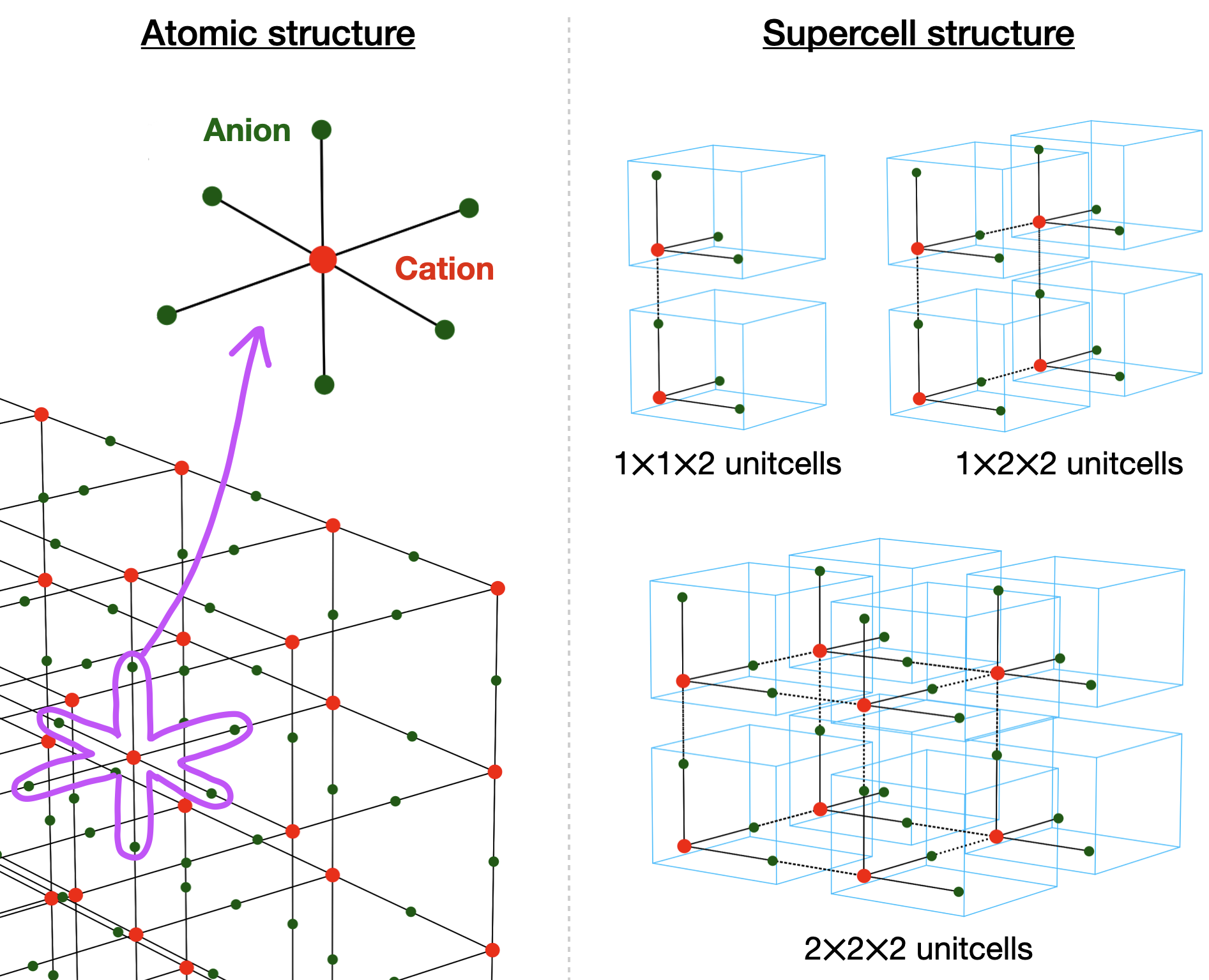}
    \caption{\small (Left) The atomic structure of MHP that is effectively described with a BX$_3$ crystalline, where each cation (B, red circle) is surrounded by six anions (X, green circles). (Right) Illustration of three different MHP supercells that are different
    in sizes, being made up of 1$\times$1$\times$2, 1$\times$2$\times$2, and 2$\times$2$\times$2 unitcells. Note that each unitcell containing 4 atoms is outlined with a sky-blue box.}
    \label{fig:perovskite}
\end{figure}

\section{Results and Discussion}\label{sec:res_disc}

Now, here we rigorously examine the feasibility of the VQE solver that incorporates the grouping strategy and measurement scheme presented in the section \ref{sec:methods_sbghz}, to find band-gap energies of 3D-confined atomic structures that are
described with a TB model. For the case problem, we consider metal-halide-perovskite (MHP) supercells that are based on the ABX$_3$ crystal structure and can flexibly generate visible lights of various colors by engineering the supercell size and the
composition of halogen atoms \cite{Richard2016, Ryu2019}. In the particular case where A and B are Methylammonium (CH$_3$NH$_3$) and Lead (Pb), respectively, the molecule A is weakly coupled and the ABX$_3$ crystal can be effectively described
with a BX$_3$ configuration as Fig.~\ref{fig:perovskite} shows, enabling an empirical representation of the corresponding electronic structure with a $sp^3$ TB model \cite{Richard2016}. In consequence, CH$_3$NH$_3$PbX$_3$ supercells of finite sizes
can be computationally constructed with repeated placements of the unitcell consisting of one cation (B(Pb)) and three anions (X) as illustrated in the right subfigures of Fig.~\ref{fig:perovskite}.

\begin{figure}[t]
    \centering
    \includegraphics[width=\columnwidth]{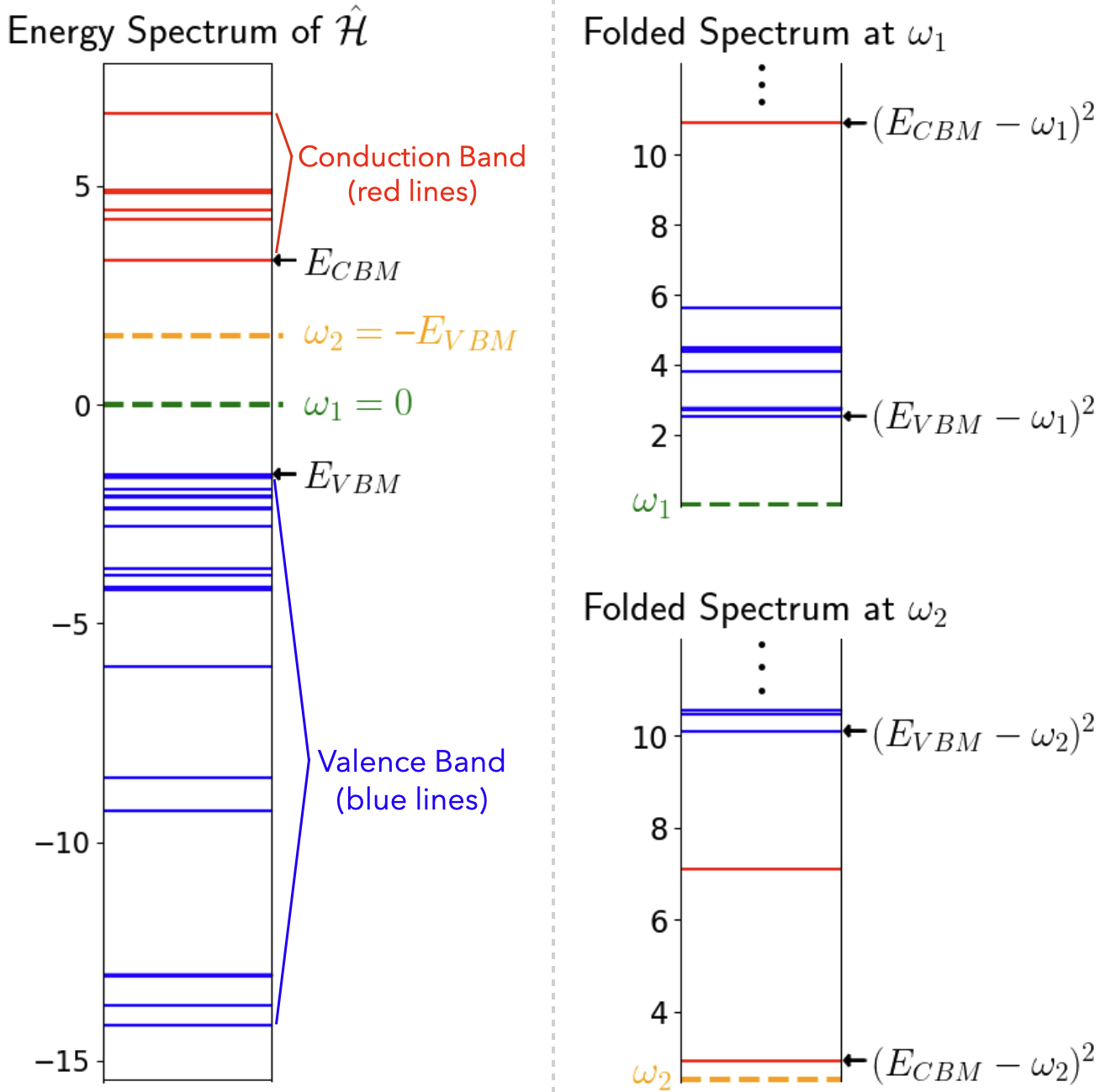}
    \caption{\small (Left) The energy spectrum of a $sp^3$ TB Hamiltonian $\hat{\mathcal{H}}$ that represents the electronic structure of a CH$_3$NH$_3$PbI$_3$ supercell consisting of 1$\times$1$\times$2 unitcells. (Right) the corresponding folded
    spectrum of $(\hat{\mathcal{H}} - \omega I)^2$ with $\omega$ = 0 (top) and $\omega$ = -$E_{VBM}$ (bottom). Results show $E_{VBM}$ and $E_{CBM}$ are located near zero, where $E_{VBM}$ is closer to zero than $E_{CBM}$.}
    \label{fig:foldedspectrum}
\end{figure}

What we aim to find is the band-gap energy of a MHP supercell, which can be obtained with the difference in energy between the minimum energy-level in conduction band and the maximum energy-level in valence band that are embedded in the middle
of the dense energy spectrum of a TB Hamiltonian as shown in the left subfigure of Fig.~\ref{fig:foldedspectrum}. Although numerous VQE studies have focused on finding excited energy states beyond the ground state of a given Hamiltonian, they mainly
rely on the fact that eigenvectors of different states are orthogonal, to seek for states in higher energy with the information of states in lower energy that are already secured \cite{Tilly.2022, Bharti.2022}. Such strategies however would not be suitable in
our case since the valance band of solid systems normally has many energy levels and we may need to compute too many redundant eigenvalues until the band-gap energy is secured. We therefore strategically employ the so-called $folded$-$spectrum$
method \cite{Peruzzo.2014}, targeting to find a specific energy level with a cost function defined as follows,

\begin{equation}\label{eq:fs}
C(\boldsymbol{\theta}) = \langle\psi(\boldsymbol{\theta})|(\hat{\mathcal{H}}-\omega I)^2 |\psi(\boldsymbol{\theta})\rangle, 
\end{equation}
where $\omega$ is the user-defined reference value around which the eigenvalue spectrum of a given Hamiltonian $\hat{\mathcal{H}}$ is folded. The cost function given in Eq.~(\ref{eq:fs}) is constructed against TB Hamiltonians with the SB grouping and
GHZ measurement, and, though the folded-spectrum method requires the circuit decomposition of a squared Hamiltonian, the SB grouping still works in a cost-efficient manner as discussed in the section \ref{sec:methods_sbghz}. Through minimization
of Eq.~\eqref{eq:fs}, we can find the square of an eigenvalue of $\hat{\mathcal{H}}$ that is closest to $\omega$, and the selection of an appropriate $\omega$ for securing the band-gap energy is straightforward since the two energy levels of our
interest -- the valence band maximum (VBM) and the conduction band minimum (CBM) -- normally lie in the vicinity of zero energy, having a negative and a positive value, respectively. Additionally, the VBM is usually closer to zero than the CBM as the
hole effective mass in solid systems is generally much larger than the electron effective mass. So, the band-gap energy can be determined with the following two steps: we first set $\omega$ to zero to find the energetic position of VBM ($E_{VBM}$) by
minimizing $C(\boldsymbol{\theta})$, and then determine the energetic position of CBM ($E_{CBM}$) by minimizing $C(\boldsymbol{\theta})$ with $\omega$ = -${E_{VBM}}$, as illustrated in the right subfigure of Fig.~\ref{fig:foldedspectrum}.

\begin{figure}[t]
    \centering
    \includegraphics[width=\columnwidth]{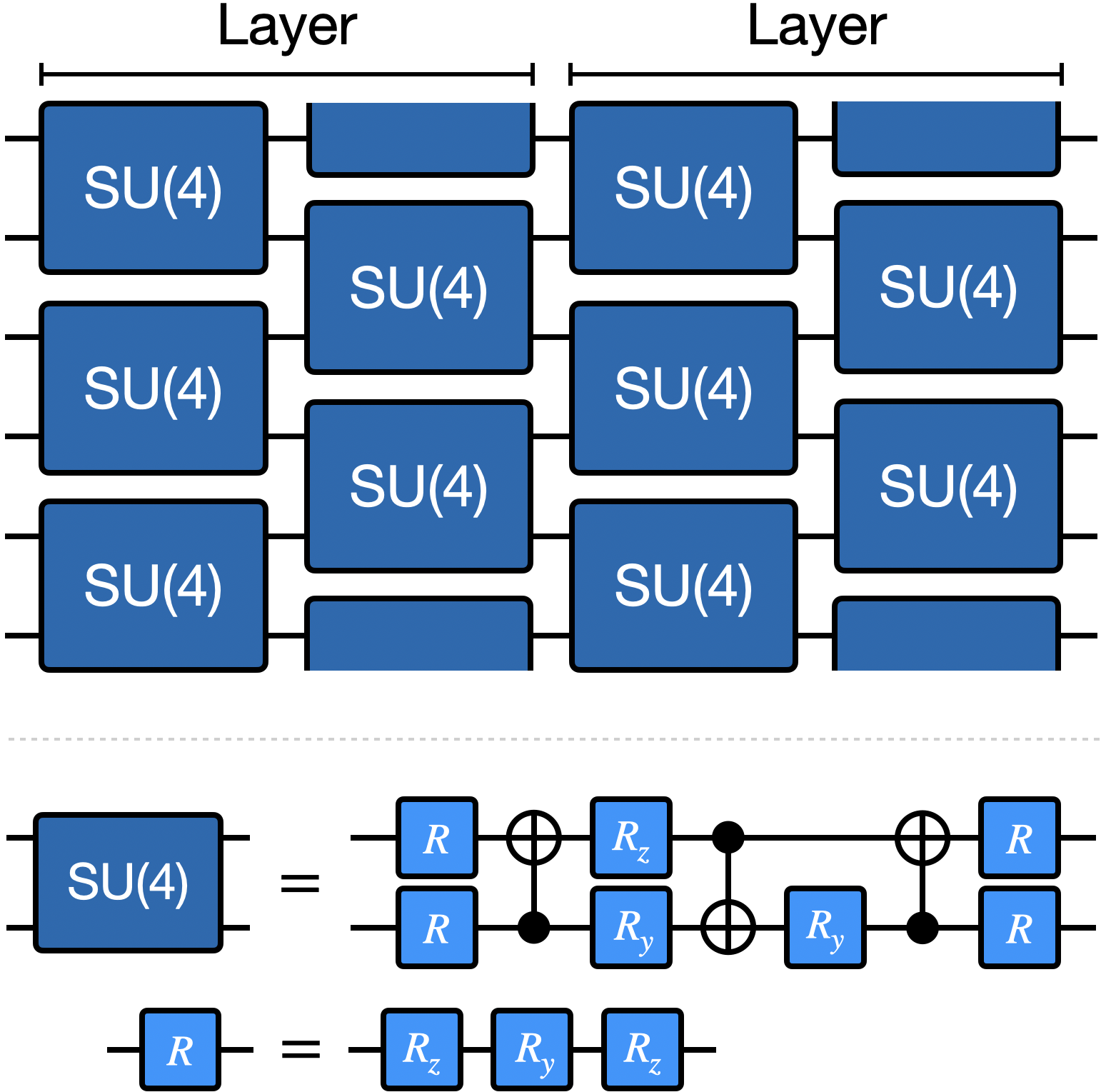}
    \caption{\small A SU(4)-based ansatz circuit in a brick-wall pattern with two layers that has been proposed by Anselmetti $et$ $al$. \cite{Anselmetti.2021}. The structure of the two-qubit logic block SU(4) is implemented based on the work of Shende
    $et$ $al.$ \cite{Shende.2004}. The optimal number of layers, which is used for actual computation, is empirically determined for each supercell whose band-gap energy is to be evaluated.}
    \label{fig:ansatz}
\end{figure}

To secure the ansatz circuit $V(\boldsymbol{\theta})$ that is realizable in NISQ devices with high customizability, which is essential to generate a variety of eigenvectors with a limited set of parameters $\boldsymbol{\theta}$, we exploit a quantum circuit
consisting of two-qubit SU(4) unitary blocks that are arranged in a brick-wall pattern \cite{Anselmetti.2021,Shende.2004} as depicted in Fig.~\ref{fig:ansatz}. The number of layers (depth) of the ansatz circuit (see Fig.~\ref{fig:ansatz} for the definition
of a single layer), which also serves as a control factor of the expressivity of $V(\boldsymbol{\theta})$ for eigenvectors, is tuned from 4 to 8 depending on the perovskite supercell to be computed. The entire workflow of a VQE calculation, including the
minimization process of the cost function given in Eq.~\ref{eq:fs}, is implemented with the PennyLane software development kit in conjunction with the ADAM optimizer and JAX library \cite{pennylane}. 

\begin{figure*}[t]
	\centering
	\includegraphics[width=\textwidth]{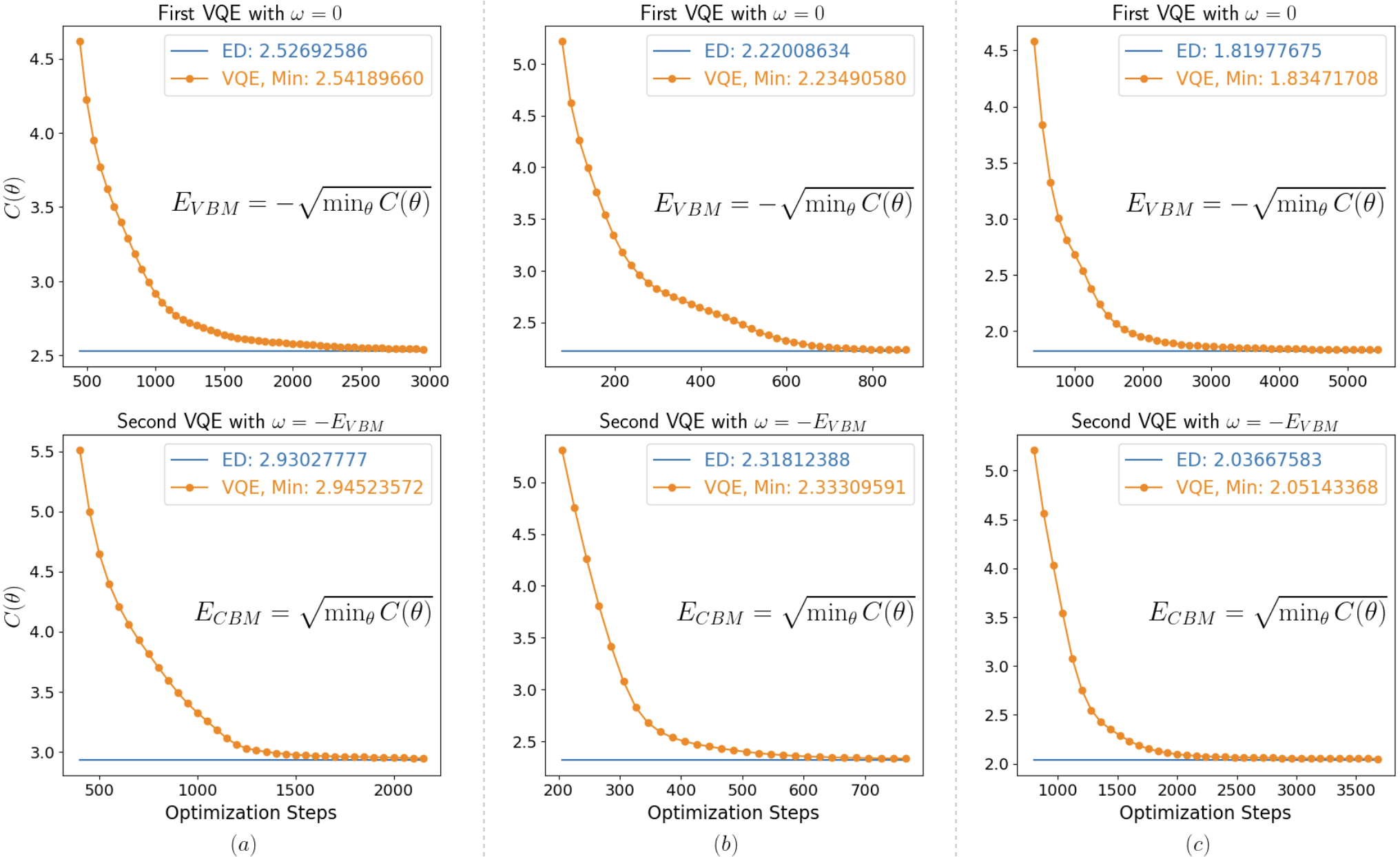}
	\caption{\small Results of VQE calculations that are conducted to find the band-gap energy of a CH$_3$NH$_3$PbI$_3$ supercell consisting of (a) 1$\times$1$\times$2 unictells, (b) 1$\times$2$\times$2 unitells, and (c) 2$\times$2$\times$2 unitcells.
	For all the cases, the cost function given in Eq.~\eqref{eq:fs} is minimized with the folded-spectrum method to obtain the minimum energy-level in conduction band ($E_{CBM}$) and the maximum energy-level in valence band ($E_{VBM}$). Optimization
	processes are conducted in an iterative manner until the cost function converges within 1.5 $\times$ $10^{-2}$. Note that the exact value of the cost function, which is calculated with the exact diagonalization (ED) method, is plotted with a blue line.}
\label{fig:res_vqe}
\end{figure*}

With the VQE scheme that has been described so far in this section and the section \ref{sec:methods_sbghz}, now we solve three finite CH$_3$NH$_3$PbI$_3$ supercells (the halogen atom X is set to the Iodine (I)) that are different in sizes, being composed
of 1$\times$1$\times$2, 1$\times$2$\times$2, and 2$\times$2$\times$2 unitcells. Since we describe atomic structures with a $sp^3$ TB model, every atom in supercell domains is represented with 4 bases ($s$, $p_x$, $p_y$, $p_z$ orbital) if no spin-orbit
coupling is included (8 bases with spin-orbit coupling). Since a single unitcell of CH$_3$NH$_3$PbI$_3$ has four atoms as depicted in Fig.~\ref{fig:perovskite}, the corresponding TB Hamiltonian matrices become 64$\times$64, 128$\times$128, and
256$\times$256, and can be represented with 7 (= 6 + 1), 8 (= 7 + 1), and 9 (= 8 + 1) qubits, respectively, as our proposed method requires one extra qubit for implementation of the GHZ measurements.

Results of VQE calculations, which are summarized in Fig.~\ref{fig:res_vqe}, demonstrates how the evaluated cost function for each supercell changes with respect to the iteration number of the optimization loop in an ideal condition where realistic effects such
as gate errors, decoherence, and finite sampling errors are not considered. The top subfigures show the first iterative processes for securing $E_{VBM}$ that are conducted with $\omega$ = 0, while the bottom ones show the second processes conducted with
$\omega$ = -$E_{VBM}$ to get $E_{CBM}$. Starting with randomly determined variational parameters, all the calculations are conducted with the ansatz circuit that has 4 (1$\times$1$\times$2 unitcells), 7 (1$\times$2$\times$2 unit cells), and 8 (2$\times$2$\times$2
unitcells) layers, and the optimization process continues until the evaluated cost function fluctuates within 1.5$\times$10$^{-2}$. When compared to the reference value of the cost function that is obtained with the exact diagonalization (ED) method (marked with
a blue line in every subfigure of Fig.~\ref{fig:res_vqe}), the overall experimental results here clearly confirm that our VQE scheme works reasonably well, indicating the solid potential for empirical simulations of electronic structures based on a TB model.

\begin{figure}[t!]
    \centering
    \includegraphics[width=\columnwidth]{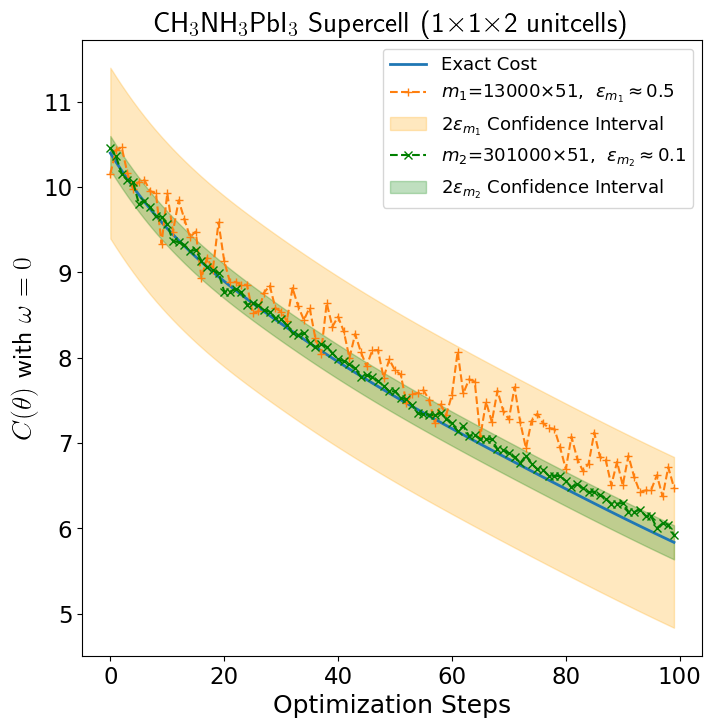}
    \caption{\small Early stage of the optimization process for VQEs in presence of finite sampling errors. The cost function targets the lowest eigenvalue of the folded spectrum of the TB Hamiltonian with $\omega$ = 0. Numerical simulations are conducted with
    two cases of sampling -- $m_1$ (orange dashed line with circles) and $m_2$ (green dashed line with crosses), and each shaded area represents the corresponding $2\epsilon$ confidence interval. The blue line corresponds to the value of the cost function
    (blue line) is obtained with no sampling errors.}
    \label{fig:vqe_shot}
\end{figure}

To examine how the working performance of our VQE scheme is affected by sampling errors that arise in reality during the measurement process, we conduct additional calculations against the CH$_3$NH$_3$PbI$_3$ supercell consisting of 1$\times$1$\times$2
unitcells. We consider two error rates, $\epsilon_1$ = 0.5 and $\epsilon_2$ = 0.1 that drive the optimal shot number of $\sim$6.15$\times$10$^5$ and $\sim$1.53$\times$10$^7$, respectively, with Eq.~\eqref{eq:samp_err_bound}. Since the cost function in this
case can be evaluated with a total of 51 circuits, we assume that every circuit needs the same number of shots, and the number of required shots per circuit becomes $\sim$1.3$\times$10$^4$ and $\sim$3.01$\times$10$^5$ for $\epsilon_1$ and $\epsilon_2$,
respectively. The early behavior of the optimization process in our VQE scheme is shown in Fig.~\ref{fig:vqe_shot}, where the value of the cost function is plotted with a $2\epsilon$ confidence interval that is centered on the error-free value. Results say that the
values under sampling errors generally reside within their confidence intervals, indicating the utility of Eq.~\eqref{eq:samp_err_bound} as a rule of thumb for determination of required shot numbers. As the optimization process evolves, however, the inaccuracy
becomes more noticeable obviously due to the accumulation of sampling errors. Apparently, the number of shots cannot be increased blindly, and thus it must be determined appropriately at the sacrifice of the precision of solutions.

\begin{figure}[htb!]
\centering
\includegraphics[width=\columnwidth]{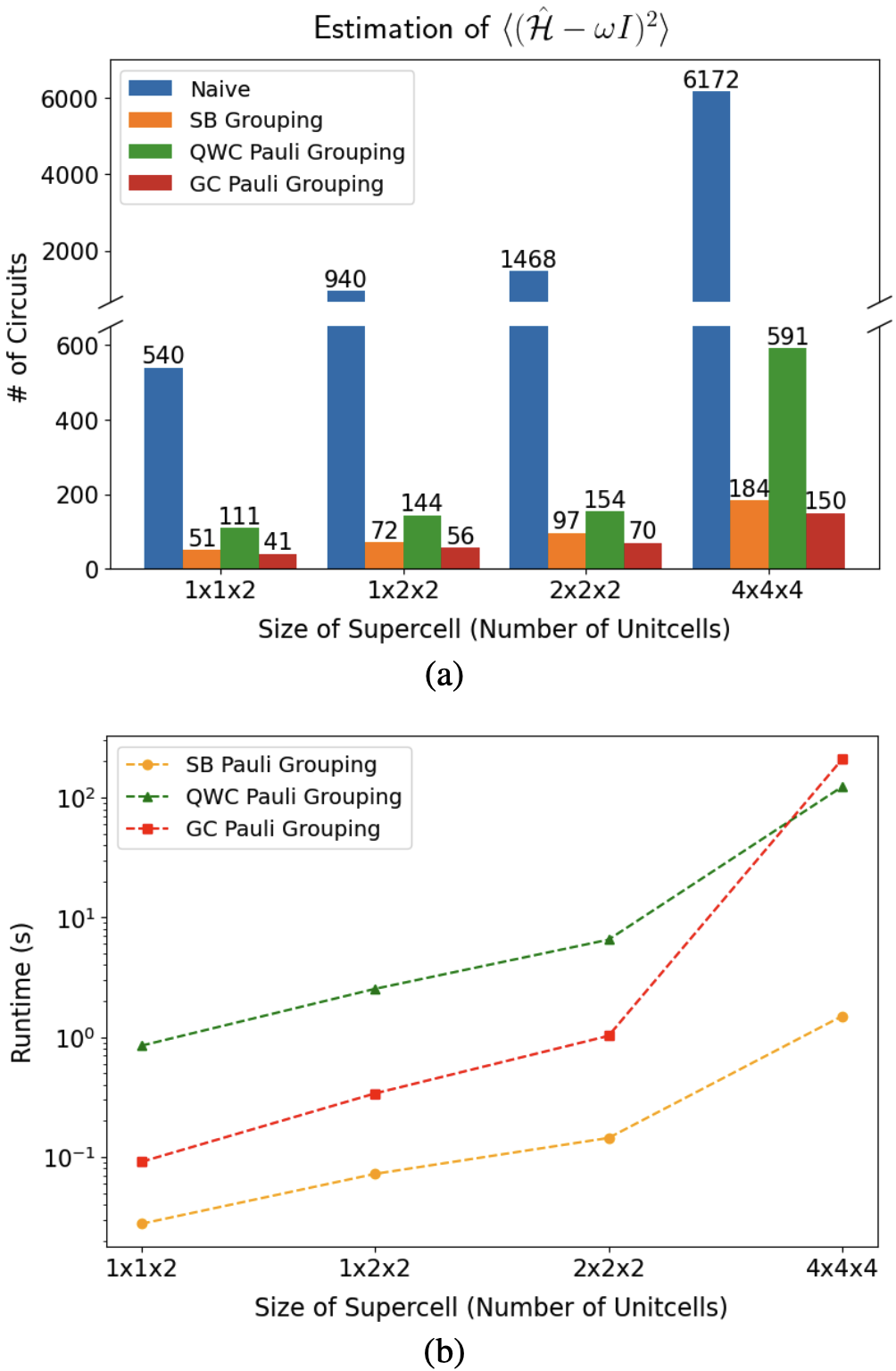}
\caption{\small Superiority of the SB grouping to the QWC and GC Pauli grouping for handling the cost function given in Eq.~\eqref{eq:fs}. (a) The number of required circuits is shown as a function of supercell sizes, where the ``Naive'' case indicates the results
secured with the raw Pauli decomposition. (b) The time required to complete the mapping process of TB Hamiltonians clearly supports the superiority of our approach based on the SB grouping to the Pauli-based grouping methods that require NP-hard optimization
processes.}
\label{fig:groupings}
\end{figure}

Though our VQE scheme cannot be free from the sampling issue as other quantum algorithms do, its strength - the cost efficiency driven with the SB grouping and the GHZ measurement - is quite obvious compared to VQEs employing other Pauli-based approaches.
To examine the efficiency of our method for TB simulations rigorously, we construct the cost function given in Eq.~\eqref{eq:fs} with various mapping strategies for Hamiltonians of CH$_3$NH$_3$PbI$_3$ supercells whose dimensions range from 1$\times$1$\times$2
(6 qubits) to 4$\times$4$\times$4 unitcells (12 qubits). Results given in Fig.~\ref{fig:groupings}(a) clearly show that, in terms of the number of quantum circuits required to construct the cost function, the performance of our approach surpasses that of the QWC Pauli
grouping method and closely approaches that of the GC Pauli grouping method. We note that these results are well aligned to the underlying grouping conditions: the measurement circuit coupled to the GC Pauli grouping method has up to $O(kN/\log k)$ CNOT gates
whereas the QWC Pauli grouping method does not require CNOT gates for measurements. The SB grouping method, which employs up to $N$ CNOT gates for measurements, would be thus positioned between the two Pauli-based methods. In spite of the similarity
in the required number of circuits, the advantage of the SB grouping over the GC Pauli grouping is still clear in terms of the computing speed as Pauli-based grouping methods must involve an optimization problem that is NP-hard \cite{Yen.2020,Verteletskyi.2020dt}, 
whereas the SB groping method just requires up to $O(M)$ runtimes as outlined in Algorithm~\ref{alg}. Fig.~\ref{fig:groupings}(b), which shows the time required to complete mapping Hamiltonians into quantum circuits in an Apple Mac Studio (M2 Max processor \&
64GB memory), clearly supports the computational advantage of our method over the GC and the QWC Pauli grouping. Though heuristic approaches are often exploited for GC Pauli grouping, they in principle cannot guarantee the quality of solutions particularly as
the problem size grows. In consequence, our method involving the SB grouping offer a more practical and scalable route for handling TB Hamiltonians, and arbitrary sparse matrices in a more broad sense.

\section{Conclusion}\label{sec:conclusion}
We have presented a variant of VQE tailored for finding the band-gap energy of the confined atomic system whose electronic structure is described with the empirical tight-binding (TB) model. Employing a set of standard basis (SB) operators to build the sparse Hamiltonian
directly from the lattice geometry of the target atomic system, the proposed VQE scheme evaluates the expectation value of the TB Hamiltonian by grouping SB operators in conjunction with the so-called Greenberger-Horne-Zeilinger (GHZ) measurement circuit that is
an extension of the Bell measurement circuit. The superiority of our method to Pauli-based approaches in terms of computational efficiency is quite obvious particularly for TB simulations, because the SB-based expression of the Hamiltonian can be done with $O(M)$
($M$ = the number of nonzero elements in the Hamiltonian) while the cost becomes $O(4^N)$ ($N$ = the qubit size of circuits) with the Pauli decomposition. Our grouping strategy coupled to the GHZ measurement evaluates the expectation value of the Hamiltonian with
fewer circuits than the case of the Pauli grouping based on the qubit-wise commutativity, and also outperforms the Pauli grouping based on the general commutativity as it involves no NP-hard optimization processes that are essential in Pauli-based methods. Results of
experiments that have been conducted to determine band-gap energies of CH$_3$NH$_3$PbI$_3$ supercells, clearly support the computational benefit of our scheme, reinforcing the potential of VQE for empirical simulations of electronic structures.

\begin{acknowledgments}
This work has been carried out under the financial support from the Korea Institute of Science and Technology Information (grant no.: P25011 and K25L1M3C3).
\end{acknowledgments}

\appendix

\section{Derivation of Eq.~\eqref{eq:prob_diff}}\label{sec:derv1}
Let us consider an $N$-qubit quantum state $|\psi\rangle$. 
In order to directly estimate the expectation value of $N$-qubit SB operators in the state $|\psi\rangle$, we prepare an ancillary qubit in the state $|G\rangle = G|0\rangle$ with $G\in\{H, S\cdot H\}$ and apply a single GHZ measurement to the composite ($N$+1)-qubit state $|G\rangle|\psi\rangle$.
The GHZ measurement is expressed as $U^\dagger_{GHZ}|z_0\mathbf{z}\rangle\langle z_0\mathbf{z}|U_{GHZ}$, where the unitary $U_{GHZ}$ is defined as Eq.~\eqref{eq:U_Bell} and $z_0\mathbf{z}$ is $z_0z_1z_2\cdots z_N$.
The probability difference $p(0\mathbf{z})$ and $p(1\mathbf{z})$ from repeated GHZ measurements yields the expectation value of a Hermitian conjugate pair of SB operators as follows:
\begin{widetext}
\begin{align}
    \nonumber p_G(0z_1\cdots z_N)-p_G(1z_1\cdots z_N) &= \langle G|\langle\psi|U^\dagger_{GHZ}\left(\hat{Z}_0 \otimes |\mathbf{z}\rangle\langle \mathbf{z}|\right)U_{GHZ}|G\rangle|\psi\rangle\\
    &= \langle G|\langle\psi|\left(\prod_{j\in\mathcal{X}}CX_{0,j}\right)\left(\hat{X}_0 \otimes  |\mathbf{z}\rangle\langle\mathbf{z}|\right)\left(\prod_{j\in\mathcal{X}}CX_{0,j}\right)|G\rangle|\psi\rangle\\
    &= \langle G|\langle\psi|\left(|0\rangle\langle 1| \otimes \left[|\mathbf{z}\rangle\langle\mathbf{z}| \prod_{j\in\mathcal{X}}\hat{X}_j\right] + h.c. \right) |G\rangle|\psi\rangle\\
    &= \langle G|\langle\psi|\left(|0\rangle\langle 1| \otimes |\mathbf{z}\rangle\langle\mathbf{z}\oplus\mathbf{x}| + h.c.\right)|G\rangle|\psi\rangle\\
    &= \langle G|0\rangle\langle 1|G\rangle\cdot\langle\psi|\mathbf{z}\rangle\langle\mathbf{z}\oplus\mathbf{x}|\psi\rangle + \langle G|1\rangle\langle 0|G\rangle\cdot\langle\psi|\mathbf{z}\oplus\mathbf{x}\rangle\langle\mathbf{z}|\psi\rangle\\
    &=\begin{dcases}
    \frac{1}{2}(\langle\psi|\mathbf{z}\rangle\langle\mathbf{z}\oplus\mathbf{x}|\psi\rangle + \langle\psi|\mathbf{z}\oplus\mathbf{x}\rangle\langle\mathbf{z}|\psi\rangle) & \text{if } G=H \\
    \frac{i}{2}(\langle\psi|\mathbf{z}\rangle\langle\mathbf{z}\oplus\mathbf{x}|\psi\rangle - \langle\psi|\mathbf{z}\oplus\mathbf{x}\rangle\langle\mathbf{z}|\psi\rangle) & \text{if } G=S\cdot H.
    \end{dcases}
\end{align}
\end{widetext}
Here, $\mathbf{x}$ is the $N$-bit string corresponding to the set $\mathcal{X}$, where $x_j=1$ if $j\in\mathcal{X}$ and $x_j=0$ otherwise. 
Thus, the configuration of the CNOT gates in $U_{GHZ}$ determines the bit string $\mathbf{x}$.
Conversely, if the expectation value of a specific SB operator $|\mathbf{z}\rangle\langle\mathbf{z}'|$ is to be estimated, we calculate $\mathbf{x}=\mathbf{z}\oplus\mathbf{z}'$ and define the set $\mathcal{X} = \{j \mid x_j=1, \text{ for } 1\le j \le N\}$. 

\section{Derivation of Eq.~\eqref{eq:samp_err_bound}}\label{sec:derv2}
Given an observable expressed in Eq.~\eqref{eq:obs_ghzm} and applying the sample variance formula, the variance of its expectation value $\text{Var}(\langle\hat{\mathcal{O}}\rangle)$ can be written as Eq.~\eqref{eq:samp_err}. 
For each bit string $\mathbf{x}$ and the corresponding $o_{\mathbf{z},\mathbf{z}\oplus\mathbf{x}}$, the GHZ measurement is performed on the composite state $|G\rangle|\psi\rangle$ with $m_{R(I), \mathbf{x}}$ shots, where the measurement configuration, specifically the set $\mathcal{X}$ in $U_{GHZ}$ and the gate $G\in\{H, S\cdot H\}$, is determined by $\mathbf{x}$ and whether the coefficient is real or imaginary, respectively.
The total number of shots in estimation of $\langle\hat{\mathcal{O}}\rangle$ is then given by
\begin{equation}\label{eq:total_shot}
m=\sum_{\mathbf{x}} m_{R, \mathbf{x}} + \sum_{\mathbf{x}\neq 0} m_{I, \mathbf{x}}.
\end{equation}
To minimize $m$ subject to the constraint Eq.~\eqref{eq:samp_err}, we utilize the method of Lagrange multipliers. 
The Lagrangian $\mathcal{L}$ is constructed as
\begin{align}
\mathcal{L} =  m+\lambda\left( \sum_{\mathbf{x}} \frac{\sigma^2_{R,\mathbf{x}}}{m_{R,\mathbf{x}}} + \sum_{\mathbf{x}\neq 0} \frac{\sigma^2_{I,\mathbf{x}}}{m_{I, \mathbf{x}}} - \epsilon^2 \right),
\end{align}
where $\sigma^{2}_{R, \mathbf{x}}=\text{Var}(\sum_{\mathbf{z}}\text{Re}[o_{\mathbf{z},\mathbf{z}\oplus\mathbf{x}}]|\mathbf{z}\rangle\langle\mathbf{z}\oplus\mathbf{x}|)$ and $\sigma^{2}_{I, \mathbf{x}}=\text{Var}(\sum_{\mathbf{z}}i\,\text{Im}[o_{\mathbf{z},\mathbf{z}\oplus\mathbf{x}}]|\mathbf{z}\rangle\langle\mathbf{z}\oplus\mathbf{x}|)$.
Taking the partial derivative of $\mathcal{L}$ with respect to $m_i$ gives:
\begin{align}
\frac{\partial \mathcal{L}}{\partial m_i} =  1 - \lambda \frac{\sigma^2_i}{m^2_i} = 0,
\end{align}
which yields the optimal solution $m^\ast_i = \sqrt{\lambda}\sigma_i$ for each $i \equiv (R(I), \mathbf{x})$.
By substituting this into Eq.~\eqref{eq:total_shot} and Eq.~\eqref{eq:samp_err}, the minimum total number of shots, denoted as $m^\ast$, can be derived as follow,
\begin{align}
m^\ast = \frac{1}{\epsilon^2} \left( \sum_\mathbf{x} \sigma_{R,\mathbf{x}} + \sum_{\mathbf{x} \neq 0} \sigma_{I,\mathbf{x}} \right)^2.
\end{align}
If the variances $\sigma^2_i$ can be evaluated for all $i$, the number of shots required to estimate $\langle\hat{\mathcal{O}}\rangle$ within a desired sampling error $\epsilon$ can be directly determined. 
Alternatively, we approximate them by deriving upper bounds, which allows for practical shot allocations without requiring prior knowledge of the quantum state.
The upper bound on the variance is derived as follows,
\begin{align}
\nonumber \sigma^2_i =\;&\text{Var}\left(\sum_{\mathbf{z}}c_{\mathbf{z},\mathbf{z}\oplus\mathbf{x}}|\mathbf{z}\rangle\langle\mathbf{z}\oplus\mathbf{x}|\right)\\
\nonumber &\le\bigg\langle\left(\sum_{\mathbf{z}}|c_{\mathbf{z},\mathbf{z}\oplus\mathbf{x}}|^2|\mathbf{z}\rangle\langle\mathbf{z}|\right)\bigg\rangle=\sum_\mathbf{z} |c_{\mathbf{z},\mathbf{z}\oplus\mathbf{x}}|^2p(\mathbf{z}) \\
\label{eq:var_ineq}& \le \max_{\mathbf{z}}|c_{\mathbf{z},\mathbf{z}\oplus\mathbf{x}}|^2,
\end{align}
where $c_{\mathbf{z},\mathbf{z}\oplus\mathbf{x}}$ denotes either $\text{Re}[o_{\mathbf{z},\mathbf{z}\oplus\mathbf{x}}]$ or $\text{Im}[o_{\mathbf{z},\mathbf{z}\oplus\mathbf{x}}]$.
This upper bound leads directly to the bound on the minimum total number of shots $m^\ast$ stated in Eq.~\eqref{eq:samp_err_bound}.

\bibliography{apssamp}

\end{document}